\documentclass{article}
\usepackage[utf8]{inputenc}
\usepackage{authblk}
\usepackage{setspace}
\usepackage[margin=1.25in]{geometry}
\usepackage{graphicx}
\graphicspath{ {./figures/} }
\usepackage{subcaption}
\usepackage{amsmath}
\usepackage{lineno}
\usepackage{braket}
\usepackage{acronym}
\usepackage{ulem}
\usepackage{xcolor}

\usepackage[style=nejm, 
citestyle=numeric-comp,
sorting=none]{biblatex}
\title{Detection of Earth's free oscillations utilizing TianQin}

\author[1,2]{Yuxin Yang}
\author[2]{Kun Liu}
\author[1,2*]{Xuefeng Zhang}
\author[1,2*]{Yi-Ming Hu}

\affil[1]{MOE Key Laboratory of TianQin Mission, TianQin Research Center for Gravitational Physics,
Frontiers Science Center for TianQin, Gravitational Wave Research Center of CNSA,
Sun Yat-sen University (Zhuhai Campus), Zhuhai, 519082, China.}
\affil[2]{School of Physics and Astronomy, Sun Yat-sen University (Zhuhai Campus), Zhuhai, 519082, China.}
\affil[*]{Address correspondence to: zhangxf38@sysu.edu.cn, huyiming@sysu.edu.cn}

\date{}

\begin{document}

\acrodef{PREM}[PREM]{Preliminary Reference Earth Model}
\acrodef{MCMC}[MCMC]{Markov Chain Monte Carlo}
\acrodef{TDI}[TDI]{time-delay interferometry}
\acrodef{GW}[GW]{gravitational wave}
\acrodef{SNR}[SNR]{signal-to-noise ratio}
\acrodef{PSD}[PSD]{power spectral density}
\acrodef{FIM}[FIM]{Fisher information matrix}

\maketitle

\begin{abstract}
    The measurement of Earth's free oscillations plays an important role in studying the Earth's large-scale structure.
    Space technology development presents a potential method to observe these normal modes by measuring inter-satellite distances.
    However, the disturbance from the Earth's low-degree gravity field makes it challenging for low Earth orbit gravity measurement satellites such as Gravity Recovery and Climate Experiment (GRACE) and TianQin-2 to extract signals from Earth's free oscillations directly.
    Here, we propose that by taking advantage of the high Earth orbit, the TianQin satellites can effectively avoid this disturbance, enabling direct measurement of Earth's free oscillations.
    We derive an analytical waveform to describe the response of Earth's free oscillations in TianQin.
    Based on this waveform, we use Bayesian analysis to extract the normal modes from numerical simulation data and perform parameter estimation.
    Our findings reveal that for a magnitude 7.9, Wenchuan-like earthquake, the resulting free oscillations will generate a signal that signal-to-noise ratio (SNR) is 73 in TianQin, and approximately 9 different modes can be distinguished.
    This result shows TianQin can open a new window to examine the Earth's free oscillations and study the Earth's interior and earthquakes independently from ground-based gravity measurement.
\end{abstract}


\section{Introduction}
Earth's free oscillations, also known as normal modes of the Earth, are standing waves along the surface and radius of the Earth.
Similar to natural vibrations, the Earth's free oscillations occur as the planet responds to disturbances like earthquakes.
The amplitudes, frequencies, and damping times of the Earth's free oscillations are governed by the intrinsic properties of the Earth.
Therefore, detailed measurement of the Earth's free oscillations provides a unique window to constrain the large-scale structure of the Earth. 
For example, the observation of normal modes provides information to describe the Earth's interior structure \cite{Dziewonski_1971_Overtones, Dziewonski_Anderson_PREM_1981}, the inner core of Earth \cite{Dziewonski_1971_solidity, Woodhouse_1986_evidence, Tromp_1993_support, Romanowicz_2000_anomalous, Masters_Jordan_Silver_Gilbert_1982, Ishii_1999_Normal}, and the source mechanism of earthquakes \cite{Gilbert_Dziewonski_1975, Ritzwoller_Lavely_1995, Park_2005_earth}.

After the 1960 Chilean earthquake, a long-period oscillation of local gravitational acceleration was reported.
The pattern closely follows the theoretical prediction of the Earth's free oscillations, and therefore it was regarded as the first observation \cite{Ness_1961_observations, Benioff_1961_excitation}.
Earth's free oscillation signals were initially identified by investigating the Fourier transform of the observational data from individual gravimeters \cite{Dziewonski_1972_Observations, Geller_1979_time}.
In order to enhance the \ac{SNR} and therefore increase the precision of estimated parameters, data stacking methods were proposed.
Such methods study observation data combined from multiple gravimeters \cite{Ding_2015_data}, including the spherical harmonic stacking (SHS) \cite{Buland_1979_Observations}, the optimal sequence estimation (OSE) \cite{Ding_2013_Search}, and the multistation experiment (MSE) \cite{Courtier_2000_Global}.
Technological advancements, notably the adoption of Superconducting Gravimeters, have significantly enhanced the detection precision of Earth's normal modes \cite{Rosat_2005_high}.
So far, a large number of Earth's normal modes along with their splitting spectra have been detected by ground-based stations \cite{Rosat_2003_First, Deuss_2013_New, Luan_2021_Progress}.

The advancement of space technology presents a potential alternative data source for detecting these oscillations.
The prospect of detecting Earth's normal modes by measuring inter-satellite distances was first suggested and performed by Ghobadi-Far et al. \cite{Ghobadi-Far_2019}.
However, no such oscillation signal was reported in the observation data of the Gravity Recovery and Climate Experiment (GRACE).
Its successor, the Gravity Recovery and Climate Experiment Follow-On Mission (GRACE-FO), has significantly enhanced measurement precision by three orders of magnitude\cite{Tapley_Bettadpur_Watkins_Reigber_GRACE_2004, Kornfeld_Arnold_Gross_Dahya_Klipstein_Gath_Bettadpur_GRACE-FO_2019}.
Yet again there was no reported oscillation signal.
The proposed space gravity measurement mission, TianQin-2, also holds considerable promise for detecting Earth's normal modes \cite{Mei_2020_TianQin, Gong_2021_Concepts}.
Satellites from these two missions operate in low Earth orbit on the order of hundreds of kilometers.
As shown in FIG. 5 of \cite{Abich_2019_In-Orbit}, for low-orbit gravity measurement mission, Earth's high-order global gravity would generate a significant signal at the low-frequency band ($<10^{-2}$ Hz).
The determination precision of the low-frequency band and low-order parameters is limited by this sensitivity aliasing effect.
The signals of Earth's free oscillations are normally buried under the global gravity signals.
Therefore, it is difficult to observe Earth's normal modes directly, which could explain the non-observation.
On the other hand, the space-borne \ac{GW} detector TianQin has the potential to observe Earth's free oscillations \cite{Liu_2023_effects}.
Unlike the aforementioned missions, TianQin satellites adopt a high Earth orbit ($\sim 10^{5}$ km), which is more sensitive to the low-frequency band and low-order parameters.
In addition, with a three-satellite six-links detection system instead of two-satellite two-links system and the high-precision inter-satellite distance measurement, TianQin will extract more information from the gravitational field.

TianQin is a space-borne GW detector with geocentric satellite orbit \cite{Luo_2016, Mei_2020_TianQin}.
It plans to launch three identical satellites orbiting at an altitude of $\sim 10^5$ km, forming a nearly equilateral triangle constellation.
TianQin detects GWs by measuring distance changes between satellites using laser interferometry with picometer-level precision.
The designed detection frequency band of TianQin is about $10^{-4} \sim 1$ Hz, covering the frequencies of prominent Earth's free oscillations \cite{Dziewonski_Anderson_PREM_1981}.
In this frequency range, neither the Earth's static gravitational field nor the free oscillations background will produce a detectable signal in TianQin \cite{Zhang_2021, Jiao_2023}.
However, major earthquakes such as the 2008 Wenchuan earthquake will leave a clear mark \cite{Liu_2023_effects}, suggesting that it is possible to identify  Earth's normal modes from TianQin data in the future.

In this work, we aim to develop data analysis tools to extract signals of Earth's free oscillations from simulated TianQin data.
For GW signals, the matched filtering method is often employed to extract signals from data \cite{Christensen_2022_Parameter}.
This method utilizes theoretical waveform templates to correlate and match observation data.
A high likelihood means the data has a strong response to the template.
In Section \ref{subsection: TDI_method}, we introduce the basic information of TianQin mission and \ac{TDI} method \cite{Tinto_1999_Cancellation, Armstrong_1999_Time, Tinto_Dhurandhar_2014}.
This method algebraically combines the signals from different arms of interferometry, which is anticipated to be used to cancel laser phase noise.
In Section \ref{subsection: EFOs-model}, we briefly introduce the Earth's free oscillations model used in this study, as well as the numerical method for generating simulated signals in TianQin based on this model.
In Section \ref{subsection: analytical_model_derivation}, we first derive the waveform of the Earth's free oscillation, combined with the detector's response in the first-generation TDI-X channels.
This can be achieved by following the analysis based on the satellite-to-satellite tracking work of GRACE \cite{Ghobadi-Far_2019} and TianQin \cite{Jiao_2023}, which calculates the acceleration changes between two satellites under the influence of the Earth's gravitational field based on Kaula's linear perturbation approach \cite{Kaula_1961, Kaula_1966}.
Finally, we perform a Bayesian analysis through \ac{MCMC} and \ac{FIM} in Section \ref{subsection: parameter_estimation_method}.
We summarize our results in Section \ref{section: results} and show our main conclusions in Section \ref{section: discussion}.

\section{Materials and Methods}\label{section: methods}
\subsection{TianQin mission and time-delay interferometry method}\label{subsection: TDI_method}

The primary scientific objective of TianQin is observing GW sources within the millihertz band, which requires extreme sensitivity \cite{Luo_2016}. 
The fundamental principle underlying GW detection entails the measurement of picometer-level distance variations between distinct free-fall test masses utilizing laser interferometry.
To mitigate measurement noise stemming from the laser interferometer, the interferometer arms must be of equal length. 
Consequently, in the initial proposal, the satellite orbits of TianQin would form an equilateral triangle around the Earth, facilitating the assumption of satellites operating in quasi-circular orbits as discussed in Section \ref{subsection: analytical_model_derivation}.
From a scientific perspective, the semi-major axis of the satellite orbits is approximately $10^5$ km, and the plane of the celestial equator of the satellites points to J0806.
The details of the parameters for the TianQin orbit can be found in Table \ref{tab: orbit-parameters}.

\begin{table}
    \centering
    \begin{tabular}{ccc}
        \hline
        Symbols & Parameters & values \\
        \hline
        $a$ & Orbit radius & $10^5$ km \\
        $e$ & Eccentricity & $0$ \\
        $i$ & Inclination & $74.5^{\circ}$ \\
        $\Omega$ & Longitude of ascending node & $211.6^{\circ}$ \\
        $GM$ & Earth's gravity constant & $3.986 \times 10^{-14} \mathrm{m}^3 / \mathrm{s}^2$ \\
        $R$ & Average Earth radius & $6.378 \times 10^{6} \mathrm{m}$  \\
        $\gamma$ & Spread angle between two satellites & $120^{\circ}$ \\
        $1/f_e$ & Earth rotation period & $86164 \mathrm{s}$ \\
        $1/f_o$ & Satellite rotation period & $314709 \mathrm{s}$ \\
        $\phi_{e0}$ & Earth Initial phase & $136.8^{\circ}$ \\
        $\phi_{o0}$ & Satellite Initial phase & $30^{\circ}$ \\
        \hline
    \end{tabular}
    \caption{The parameters designed for TianQin and setting in the analytical model.}
    \label{tab: orbit-parameters}
\end{table}

The design of an equal-arm interferometer for GW detectors can mitigate measurement noise originating from the laser interferometer, particularly laser frequency fluctuation noise.
Nonetheless, maintaining equal arm lengths in interferometry with sufficient precision poses a challenge for space-borne GW detectors.
Tinto et al. \cite{Tinto_1999_Cancellation} proposed the TDI method to address this issue.
This method algebraically combines the signal from different arms of interferometry, constructing equivalent arms with equal lengths to eliminate laser frequency fluctuation noise.
In this article, we focus on the first-generation TDI and provide the construction method of the TDI channel in the following.

Three symmetrical TDI channels denoted as X, Y, and Z, respectively, can be constructed for the space-borne GW detectors with satellites orbiting in an equilateral triangular configuration like TianQin.
Without loss of generality, we will only discuss the construction method for the TDI-X channel.
We denote the laser phase signals observed in satellite 1 come from different arms as $\phi_{1}(t)$ and $\phi_{1'}(t)$, which could be written as
\begin{equation}
    \begin{aligned}
        \phi_{1}(t) &= p(t-2L_3/c) - p(t) + s_{1}^{h}(t) + n_{1}(t), \\
        \phi_{1'}(t) &= p(t-2L_2/c) - p(t) + s_{1'}^{h}(t) + n_{1'}(t),
    \end{aligned}
    \label{eq: laser-phase-signal}
\end{equation}
where $p(t)$ represents the laser frequency noise, $L_{i}$ denotes the length of different arms, c denotes the speed of light, $s_{i}^{h}(t)$ are the GW signals and $n_{i}(t)$ represents other noise. 
Obviously, when the length of different arms is unequal (i.e., $L_2 \neq L_3$), the laser frequency noise persists if we construct the Michelson channel using $\Delta \phi(t) = \phi_{1}(t) - \phi_{1'}(t)$, as typically done with a laser interferometer.
Therefore, we propose defining the TDI-X channel as
\begin{equation}
    \phi_X(t) = \phi_{1}(t) - \phi_{1'}(t) - [\phi_{1}(t-2L_2/c) - \phi_{1'}(t-2L_3/c)].
    \label{eq: TDI-X-laser-phase-signal}
\end{equation}

We introduce the Time-Delay operator
\begin{equation}
    \mathcal{D}_{i} \phi(t) \equiv \phi(t-L_i/c),
\end{equation}
which satisfies
\begin{equation}
    \mathcal{D}_{i} \mathcal{D}_{j} \phi(t) = \phi(t-L_i/c-L_j/c) = \mathcal{D}_{j} \mathcal{D}_{i} \phi(t).
\end{equation}
Keeping only the laser frequency noise, the data in the TDI-X channel could be written as 
\begin{equation}
    \begin{aligned}
        \phi_X(t) &= \phi_{1}(t) - \phi_{1'}(t) - [\phi_{1}(t-2L_2/c) - \phi_{1'}(t-2L_3/c)] \\
        &= (1-\mathcal{D}_{2}^{2}) \phi_{1}(t) - (1-\mathcal{D}_{3}^{2}) \phi_{1'}(t) \\
        &= [(1-\mathcal{D}_{2}^{2})(\mathcal{D}_{3}^{2}-1)-(1-\mathcal{D}_{3}^{2})(\mathcal{D}_{3}^{2})(\mathcal{D}_{2}^{2}-1)] p(t) \\
        &= 0,
    \end{aligned}
\end{equation}
which proves that the \ac{TDI} method could cancel the laser frequency noise.
Utilizing the same method, the TDI-Y and TDI-Z channel, observed in satellites 2 and 3, respectively, could be defined as
\begin{equation}
    \phi_Y(t) = \phi_{2}(t) - \phi_{2'}(t) - [\phi_{2}(t-2L_3/c) - \phi_{2'}(t-2L_1/c)],
\end{equation}

\begin{equation}
    \phi_Z(t) = \phi_{3}(t) - \phi_{3'}(t) - [\phi_{3}(t-2L_1/c) - \phi_{3'}(t-2L_2/c)].
\end{equation}
Another widely used TDI channel, AET channel \cite{Armstrong_1999_Time}, can be defined as linear combinations of XYZ channel:
\begin{equation}
    A = \frac{1}{\sqrt{2}} (Z-X),
\end{equation}

\begin{equation}
    E = \frac{1}{\sqrt{6}} (X-2Y+Z),
\end{equation}

\begin{equation}
    T = \frac{1}{\sqrt{3}} (X+Y+Z).
\end{equation}

This kind of channel consists of three orthogonal channels, making noise uncorrelated in different channels.
A and E channels still contain GW signals, while GW response is very weak in the T channel.
Therefore, the T channel is usually used to monitor noise in GW detection.
The following derivation will focus on the Earth's free oscillations' signals in the TDI-X channel.
It's easy to generalize our method to the response on the other channels.

\subsection{Earth's free oscillations model and numerical simulation}\label{subsection: EFOs-model}

The Earth's free oscillations are categorized into spheroidal and toroidal modes \cite{Woodhouse_2010_theory}.
Toroidal modes do not induce radial deformation or alter the Earth's gravitational field. 
Conversely, spheroidal modes encompass both radial and horizontal movements, resulting in alterations to the Earth's gravitational field.
Only the latter can perturb satellite orbits, rendering them detectable by TianQin.
Herein, we exclusively focus on the spheroidal modes incited by the major earthquake.

An earthquake is commonly characterized using the point-source double-couple model along with pertinent parameters including scalar seismic moment, dip, rake, and strike \cite{Kanamori_1974}.
The amplitudes of the Earth's free oscillations, which are induced by seismic events, are determined by these earthquake parameters.
The normal modes of Earth can be derived through computation utilizing the \ac{PREM}, a widely adopted nonrotating, elastic, spherically symmetric, one-dimensional Earth model \cite{Dziewonski_Anderson_PREM_1981}.
Transforming seismic parameters into Earth's free oscillation parameters can be achieved by spherical harmonic decomposition and coordinate transformation \cite{stein_1977, Kanamori_1974, stein_2009_introduction}.
Finally, the perturbation in gravity resulting from free oscillations is expressed through a series of coefficients denoted as ${}_{l} \Delta \bar{C}_{n m}$ and ${}_{l}\Delta \bar{S}_{n m}$.

Then we introduce the details of these parameters describing the variations in Earth's gravitational field.
Within the framework of the International Terrestrial Reference System (ITRS), a satellite that is located at spherical coordinates radius $r$, co-latitude $\theta$, and longitude $\lambda$ will experience the Earth's gravity field described as
\begin{equation}
    \begin{aligned}
        & V(r, \theta, \lambda) = \frac{GM}{R} \sum_{n=0}^{N} \sum_{m=0}^{n} \left( \frac{R}{r} \right)^{n+1} \\
        & \times \left( \bar{C}_{nm} \cos(m \lambda) + \bar{S}_{mn} \sin(m 
        \lambda) \right) \bar{P}_{nm}(\cos \theta),
    \end{aligned}
    \label{eq: earth-gravity-field}
\end{equation}
where $G$ is the gravitational constant, $M$ is the Earth's total mass, $R$ is the Earth's average radius, $n$ and $m$ correspond to the degree and order of spherical harmonic expansion, respectively \cite{Montenbruck_2000}.
The symbol $N$ denotes the truncation degree of the spherical harmonics function.

Eq. \eqref{eq: earth-gravity-field} delineates a scenario wherein the gravitational field remains temporally invariant.
Upon the occurrence of a significant earthquake, resulting in the generation of Earth's free oscillations, the coefficients $\bar{C}_{nm}$ and $\bar{S}_{mn}$ cease to remain constant.
The fluctuations in the Earth's gravitational field can be modeled as \cite{Ghobadi-Far_2019,Han_Riva_Sauber_Okal_2013} 
\begin{equation}
    \left\{
        \begin{array}{l}
        \bar{C}_{n m}(t)=\bar{C}_{n m}+\sum_{l=0}^L {}_l \Delta \bar{C}_{n m} \,{}_l \xi_n\left(t-t_0\right), \\
        \bar{S}_{n m}(t)=\bar{S}_{n m}+\sum_{l=0}^L {}_l \Delta \bar{S}_{n m} \,{}_l \xi_n\left(t-t_0\right),
        \end{array}
    \right.
    \label{eq: variations-gravity-field-coefficients}
\end{equation}

where $l$ marks free oscillation overtones, $L$ represents the maximum overtone, and $t_0$ denotes the time of earthquake occurrence. 
The coefficients ${}_{l} \Delta \bar{C}_{n m}$ and ${}_{l}\Delta \bar{S}_{n m}$ describe the magnitudes of free oscillation across various spherical harmonics.
The time-dependent functions are defined as 
\begin{equation}
    { }_l \xi_n(t)= \begin{cases}0, & t<0 \\ 1-{ }_l \tau_n(t) \cos \left(2 \pi_l f_n t\right), & t \geq 0\end{cases} 
    \label{eq: variations-gravity-field-describe-1}
\end{equation}
where
\begin{equation}
    { }_l \tau_n(t)=\exp \left(-\frac{2 \pi_l f_n\left(t-t_0\right)}{2_l Q_n}\right).
    \label{eq: variations-gravity-field-describe-2}
\end{equation}

For each degree and overtones, ${}_l f_{n}$ represents the frequency of oscillations, while ${}_l Q_{n}$ denotes the attenuation factor characterizing the decay rate of free oscillations.
The frequency and attenuation factor of each oscillation overtone can be computed utilizing \ac{PREM}.

Based on the Earth's free oscillations model, a simulation of observation data can be produced by numerical method.
In this work, we use the orbit evolution simulation program \textbf{TQPOP} \cite{Zhang_2021} to generate satellites' positions numerically.
Then, we calculate the laser propagation time between satellites to produce the observation data in single links, which are subsequently transformed into TDI channel data.

However, the numerical method requires a long computation time, making the use of this waveform for signal extraction difficult.
Therefore, in this work, we propose an analytical waveform for signal extraction.
This waveform is derived based on a simplified model that assumes the TianQin satellites operate in quasi-circular orbit, with the same orbital altitude, and forms an equilateral triangle constellation.
It can describe the response of Earth's free oscillation in the first-generation TDI-X channel.
The details of the waveform derivation will be presented in Section \ref{subsection: analytical_model_derivation}.

\subsection{Analytical model derivation}\label{subsection: analytical_model_derivation}


This section aims to derive an analytical formula to describe the response of Earth's free oscillations. 
The Earth's normal modes could be modeled by the Eq. \eqref{eq: earth-gravity-field} - Eq. \eqref{eq: variations-gravity-field-describe-2}.

Firstly, we deduce the movement of satellites under the Earth's gravity field.
In the International Celestial Reference System (ICRS), the gravitational potential experienced by satellites due to Earth can be expressed as
\begin{equation}
    \begin{aligned}
        & V(a, i, e, \omega, M, \Omega, \Theta)=\frac{G M}{R} \sum_{n=0}^N\left(\frac{R}{a}\right)^{n+1} \\
        & \times \sum_{m=0}^n \sum_{p=0}^n \bar{F}_{n m p}(i) \sum_{q=-\infty}^{\infty} G_{n p q}(e) S_{n m p q}(\omega, M, \Omega, \Theta),
    \end{aligned}
    \label{eq: gravitational-potential-with-eccentricity}
\end{equation}
where the arguments $\{a,i,e,\omega,M,\Omega,\Theta \}$ represent the semi-major axis, the inclination, the eccentricity, the argument of perigee, the mean anomaly, the right ascension of the ascending node, and the Greenwich Apparent Sidereal Time (GAST), respectively. 
The inclination function $\bar{F}_{n m p}(i)$ describes the impact of satellite orbital inclination and will be discussed extensively later \cite{Kaula_1961, Kaula_1966}. The function $G_{n p q}(e)$ characterizes the impact of the eccentricity.
However, we do not address this function in the present work, as the rationale will be explained subsequently.
The function $S_{n m p q}$ is defined as 
\begin{equation}
    S_{n m p q}(\omega, M, \Omega, \Theta) = \alpha_{n m} \cos{\phi_{n m p q}} + \beta_{n m} \sin{\phi_{n m p q}}.
    \label{eq: gravitational-potential-with-eccentricity-s}
\end{equation}
The coefficients $\{\alpha_{n m}, \beta_{n m}\}$ are rearrangements of $\{\bar{C}_{n m}, \bar{S}_{n m}\}$ (Eq. \ref{eq: earth-gravity-field}),
\begin{equation}
    \begin{gathered}
        \alpha_{n m}=\left\{\begin{array}{rr}
        \bar{C}_{n m}, & n-m=\text { even } \\
        -\bar{S}_{n m}, & n-m=\text { odd }
        \end{array}\right. \\
        \beta_{n m}= \begin{cases}\bar{S}_{n m}, & n-m=\text { even } \\
        \bar{C}_{n m}, & n-m=\text { odd }\end{cases}
    \end{gathered}
    \label{eq: coefficients-rearrangements}
\end{equation}
and the angular argument reads 
\begin{equation}
    \phi_{n m p q} = (n-2p) \omega + (n-2p+q)M + m(\Omega - \Theta).
    \label{eq: gravitational-potential-with-eccentricity-phi}
\end{equation}

TianQin's satellites orbit in quasi-circular paths, enabling us to assume an eccentricity of zero for the orbits.
Therefore, we consider only the case where $e=0$ for the function $G_{npq}(e)$.
The function $G_{npq}(0)$ equals $1$ when $q=0$ and becomes zero for other values of $q$. 
Consequently, we can disregard this function in Eq. \eqref{eq: gravitational-potential-with-eccentricity}, simplifying it to \cite{Sharifi_2006}
\begin{equation}
    V=\sum_{n=0}^N \sum_{m=0}^n \sum_{k=-n[2]}^n u_{n}(a) \bar{F}^{k}_{m p}(i) (\alpha_{n m} \cos{\varphi_{mk}} + \beta_{n m} \sin{\varphi_{mk}}),
    \label{eq: gravitational-potential}
\end{equation}
with the angular argument
\begin{equation}
    \varphi_{mk} = k \varphi^o + m \varphi_e,
    \label{eq: gravitational-potential-varphi}
\end{equation}
and
\begin{equation}
    \varphi^o = \omega + M, \qquad \varphi_e = \Omega - \Theta.
\end{equation}
To be convenient, we rearrange the summation indices with $k=n-2p$ and define 
\begin{equation}
    u_{n}(a) = \frac{G M}{R} \left(\frac{R}{a}\right)^{n+1}.
\end{equation}
Then we give the expression for the inclination function 
\begin{equation}
    \begin{aligned} 
        \bar{F}_{n m}^k(i) & =\sum_{s=0}^m \sum_{t=t_1}^{t_2} \sum_{b=b_1}^{b_2} \sqrt{\left(2-\delta_{0 m}\right)(2 n+1) \frac{(n-m) !}{(n+m) !}}\left[\frac{(-1)^{b+g}}{2^{2 n-2 t}} \frac{(2 n-2 t) !}{t !(n-t) !(n-m-2 t) !}\right. \\
        & \left.\times C_m^s C_{n-m-2 t+s}^b C_{m-s}^{\frac{n}{2}-\frac{k}{2}-t-b} \sin ^{n-m-2 t}(i) \cos ^s(i)\right]
    \end{aligned}
    \label{eq: inclination-function-expression}
\end{equation}
by using the auxiliary index
\begin{equation}
    g= \begin{cases}\frac{n-m}{2}, & n-m=\text { even } \\ \frac{n-m-1}{2}, & n-m=\text { odd }\end{cases}
\end{equation}
with
\begin{equation}
    t_1=0, \quad t_2= \begin{cases}-\frac{1}{2} k+\frac{1}{2} n, & k \geq n-2 g \\ g, & k \leq n-2 g\end{cases}
\end{equation}
and
\begin{equation}
    \begin{aligned}
    & b_1=\left\{\begin{array}{cl}
    0, & 2 s-k-2 t \leq 2 m-n, \\
    s-\frac{1}{2} k-t+\frac{1}{2} n-m, & 2 s-k-2 t \geq 2 m-n,
    \end{array}\right. \\
    & b_2=\left\{\begin{array}{c}
    s-2 t+n-m, 2 s+k-2 t \leq 2 m-n, \\
    -\frac{1}{2} k-t+\frac{1}{2} n, \quad 2 s+k-2 t \geq 2 m-n .
    \end{array}\right. \\
    &
    \end{aligned}
\end{equation}

Utilizing the formula describing the gravitational potential experienced by satellites, we can calculate the range acceleration between two TianQin satellites.
Here, subscripts $\{1, 2\}$ denote two different satellites, SC1 and SC2, respectively.
Subsequently, we define $L$ as the distance between two TianQin satellites, allowing us to express the range acceleration $\ddot{L}$ as (refer to \cite{Sharifi_2006} for further details)
\begin{equation}
    \ddot{L}=\frac{1}{a}\left(\frac{\partial V_2}{\partial \varphi_2^o}-\frac{\partial V_1}{\partial \varphi_1^o}\right) \cos \left(\frac{\gamma}{2}\right)+\left(\frac{\partial V_2}{\partial r_2}+\frac{\partial V_1}{\partial r_1}\right) \sin \left(\frac{\gamma}{2}\right).
    \label{eq: range-acceleration-1}
\end{equation}

We incorporate the gravitational potential formula given by Eq. \eqref{eq: gravitational-potential} into Eq. \eqref{eq: range-acceleration-1}.
Subsequently, we substitute $\varphi^{o}$ with $\varphi^{om}$, where $\varphi^{om}$ represents the argument of latitude of the midpoint between SC1 and SC2.
Thus, the formula for range acceleration becomes \cite{Jiao_2023}
\begin{equation}
    \ddot{L} = \sum_{l=0}^{L} \sum_{n=0}^N \sum_{m=0}^n \sum_{k=-n[2]}^n \Xi_{nmk}(a, i, \gamma) [\alpha_{nm} \cos(m \varphi_e + k \varphi^{om})+\beta_{nm} \sin(m \varphi_e + k \varphi^{om})],
    \label{eq: range-acceleration-2}
\end{equation}
where
\begin{equation}
    \begin{aligned}
    & \Xi_{n m k}(a, i, \gamma)=-2 \frac{u_n(a)}{a} \bar{F}_{n m}^k(i) \\
    & \times\left[k \sin \left(k \frac{\gamma}{2}\right) \cos \left(\frac{\gamma}{2}\right)+(n+1) \sin \left(\frac{\gamma}{2}\right) \cos \left(k \frac{\gamma}{2}\right)\right].
    \end{aligned}
\end{equation}

It is important to note that the formulas derived above only account for a static gravity field.
In the scenario of a time-varying gravity field induced by the Earth's free oscillations, it is necessary to convert the static coefficients ${\bar{C}_{nm}, \bar{S}_{mn}}$ to time-varying coefficients ${\bar{C}_{nm}(t), \bar{S}_{mn}(t)}$ using Eq. \eqref{eq: variations-gravity-field-coefficients}.
Additionally, the parameters $\varphi^{om}$ and $\varphi_{e}$ are time-varying and given by $2\pi f_o t + \phi_{o0}$ and $2\pi f_e t + \phi_{e0}$, respectively, where $f_o$ represents the frequency of the satellites' rotation and $f_e$ denotes the frequency of the Earth's rotation.
Subsequently, we express the time-varying range acceleration as
\begin{equation}
    \begin{aligned}
        \ddot{L}(t) &= \sum_{n=0}^N \sum_{m=0}^n \sum_{k=-n[2]}^n \frac{1}{2} \Xi_{mnk}(a, i, \gamma) \exp \left[{}_l \lambda_{n} (t-t_0) \right] \times \\
        & [-{ }_l\Delta \alpha_{nm} \cos({}_l \omega_{mnk}^{+} (t-t_0) + \phi_{mk}) + { }_l\Delta \beta_{nm} \sin({}_l \omega_{mnk}^{+} (t-t_0) + \phi_{mk}) \\
        &-{ }_l\Delta \alpha_{nm} \cos({}_l \omega_{mnk}^{-} (t-t_0) - \phi_{mk}) - { }_l\Delta \beta_{nm} \sin({}_l \omega_{mnk}^{-} (t-t_0) - \phi_{mk})],
    \end{aligned}
    \label{eq: range-acceleration-time-varying}
\end{equation}
where 
\begin{equation}
    {}_l \lambda_{n} = \frac{2 \pi {}_l f_n}{2 {}_l Q_n},
\end{equation}
and
\begin{equation}
    {}_l \omega_{mnk}^{+} = 2 \pi \left({}_l f_n + m f_e - k f_o \right), \qquad
    {}_l \omega_{mnk}^{-} = 2 \pi \left({}_l f_n - m f_e + k f_o \right).
    \label{eq: omega-relationship}
\end{equation}
The phase of different modes can be expressed as 
\begin{equation}
    \phi_{mk} = m \phi_{e0} - k \phi_{o0}
    \label{eq: phase-earth-orbit}
\end{equation}

Upon performing two integrations, the range acceleration transforms into the differential arm length $\Delta L$.
Typically, observation data from TianQin is represented in the form of strain, denoted as $h = \Delta L / L$, where $L$ signifies the unperturbed arm length.
Thus, the observation data from a single arm, induced by Earth's free oscillations, can be formulated as follows
\begin{equation}
    \begin{aligned}
        h(t) &= \sum_{l=0}^{L} \sum_{n=0}^N \sum_{m=0}^n \sum_{k=-n[2]}^n \frac{1}{2L} \Xi_{mnk}(a, i, \gamma) \exp \left({}_l \lambda_{n} (t-t_0) \right) \times \\
        & [{}_l L_{nmk}^{+c} \cos({}_l \omega_{mnk}^{+} (t-t_0) + \phi_{mk}) + {}_l L_{nmk}^{+s} \sin({}_l \omega_{mnk}^{+} (t-t_0) + \phi_{mk}) \\
        &+{}_l L_{nmk}^{-c} \cos({}_l \omega_{mnk}^{-} (t-t_0) - \phi_{mk}) + {}_l L_{nmk}^{-s} \sin({}_l \omega_{mnk}^{-} (t-t_0) - \phi_{mk})],
    \end{aligned}
    \label{eq: strain-single-arm}
\end{equation}
where
\begin{equation}
    \begin{aligned}
        { }_l L_{mnk}^{+c} =& \frac{1}{\left(({ }_l \lambda_{n})^2 + ({ }_l \omega_{mnk}^{+})^2 \right)^2} (-{ }_l\Delta \alpha_{nm} ({ }_l \lambda_{n}^2 - ({ }_l \omega_{mnk}^{+})^2) - 2 { }_l\Delta \beta_{nm} { }_l \lambda_{n} { }_l \omega_{mnk}^{+}), \\
        { }_l L_{mnk}^{+s} =& \frac{1}{\left(({ }_l \lambda_{n})^2 + ({ }_l \omega_{mnk}^{+})^2 \right)^2} (-{ }_l\Delta \beta_{nm} ({ }_l \lambda_{n}^2 - ({ }_l \omega_{mnk}^{+})^2) + 2 { }_l\Delta \alpha_{nm} { }_l \lambda_{n} { }_l \omega_{mnk}^{+}), \\
        { }_l L_{mnk}^{-c} =& \frac{1}{\left(({ }_l \lambda_{n})^2 + ({ }_l \omega_{mnk}^{-})^2 \right)^2} (-{ }_l\Delta \alpha_{nm} ({ }_l \lambda_{n}^2 - ({ }_l \omega_{mnk}^{-})^2) + 2 { }_l\Delta \beta_{nm} { }_l \lambda_{n} { }_l \omega_{mnk}^{-}), \\
        { }_l L_{mnk}^{-s} =& \frac{1}{\left(({ }_l \lambda_{n})^2 + ({ }_l \omega_{mnk}^{-})^2 \right)^2} ({ }_l\Delta \beta_{nm} ({ }_l \lambda_{n}^2 - ({ }_l \omega_{mnk}^{-})^2) + 2 { }_l\Delta \alpha_{nm} { }_l \lambda_{n} { }_l \omega_{mnk}^{-}).
    \end{aligned}
\end{equation}

As previously mentioned, TDI is extensively employed in space-borne GW detectors.
The final data acquired from TianQin is processed by TDI.
It is necessary to derive the formula for the Earth's free oscillation signals after undergoing TDI processing.
As demonstrated in Section \ref{subsection: TDI_method}, TDI fundamentally involves algebraic calculations of data from various arms.
Consequently, we must consider the effects on different arms simultaneously.
Eq. \eqref{eq: phase-earth-orbit} reveals that the signal phase is dependent on both the Earth's phase and the orbit's phase.
The Earth's phase is the same for all arms of TianQin, while the orbit's phase is different.
Let $\phi_{mk}^{(1)}$ denote the signal phase in one arm. Then, the phase of another arm in the opposite direction of the satellites' movement can be expressed as
\begin{equation}
    \phi_{mk}^{(2)} = \phi_{mk}^{(1)} + k \gamma.
    \label{eq: phase-different-arm}
\end{equation}

Given the phase relationship between different arms as described by Equation \eqref{eq: phase-different-arm}, let $h(t, \phi_{mk})$ represent the signal in one arm, and denote the signal in another arm as $h(t, \phi_{mk}+k \gamma)$.
Recall the algebraic construction of the TDI-X channel using Eq. \eqref{eq: TDI-X-laser-phase-signal}.
Subsequently, we utilize a linear approximation to describe the time delay effect and provide the expression of the signal in the TDI-X channel
\begin{equation}
    \begin{aligned}
        h_X(t) =& \ [h(t, \phi_{mk}) + h(t-\Delta t, \phi_{mk}) + h(t-2\Delta t, \phi_{mk}+ k\gamma) + h(t-3\Delta t, \phi_{mk}+k\gamma)] \\
        & - [h(t, \phi_{mk}+k\gamma) + h(t-\Delta t, \phi_{mk}+k\gamma) + h(t-2\Delta t, \phi_{mk}) + h(t-3\Delta t, \phi_{mk})] \\
        \approx & \ 4h'(t, \phi_{mk}) \Delta t - 4h'(t, \phi_{mk}+k\gamma) \Delta t,
    \end{aligned}
    \label{eq: TDI-X-channel}
\end{equation}
where $\Delta t$ represents the laser propagation time from one satellite to another.
It is evident that the ratio of $\Delta t$ to $L$ yields the speed of light, denoted as $c$. 
Therefore, we substitute the derivative of Eq. \eqref{eq: strain-single-arm} into Eq. \eqref{eq: TDI-X-channel}, replacing $\Delta t$ and $L$ with $c$.
Finally, we obtain the result that describes the response generated by Earth's free oscillations in the TDI-X channel for TianQin.
For convenience, we revert $\varphi^{om}$, which represents the argument of latitude of the midpoint of SC1 and SC2, back to $\varphi^{o}$, which represents the argument of latitude of SC1. 
Subsequently, we present the final result as
\begin{equation}
    \begin{aligned}
        h_X(t) &= \sum_{l=0}^{L} \sum_{n=0}^N \sum_{m=0}^n \sum_{k=-n[2]}^n \exp[-{ }_l \lambda_{mn} (t-t_0)] \sum_{k=-n[2]}^{n} \frac{1}{2} \Xi_{mnk}(a, i, \gamma) \times \\
        &[{ }_l P_{nm}^{+c} \cos({ }_l \omega_{mnk}^{+} (t-t_0) + \phi_{mk}) + { }_l P_{nm}^{+s} \sin({ }_l \omega_{mnk}^{+} (t-t_0) + \phi_{mk}) \\
        &+{ }_l P_{nm}^{-c} \cos({ }_l \omega_{mnk}^{-} (t-t_0) - \phi_{mk}) + { }_l P_{nm}^{-s} \sin({ }_l \omega_{mnk}^{-} (t-t_0) - \phi_{mk})],
    \end{aligned}
    \label{eq: TDI-X-channel-final-result}
\end{equation}
where
\begin{equation}
    \begin{aligned}
        { }_l P_{mnk}^{+c} =& \frac{8}{c} \cdot \sin(\frac{k \gamma}{2}) \cdot \frac{1}{({ }_l \lambda_{n})^2 + ({ }_l \omega_{mnk}^{+})^2 } \left( \omega_{mnk}^{+} { }_l\Delta \alpha_{nm} + { }_l \lambda_{n} { }_l\Delta \beta_{nm} \right) \\
        { }_l P_{mnk}^{+s} =& \frac{8}{c} \cdot \sin(\frac{k \gamma}{2}) \cdot \frac{1}{({ }_l \lambda_{n})^2 + ({ }_l \omega_{mnk}^{+})^2 } \left( -\omega_{mnk}^{+} { }_l\Delta \beta_{nm} + { }_l \lambda_{n} { }_l\Delta \alpha_{nm} \right) \\
        { }_l P_{mnk}^{-c} =& -\frac{8}{c} \cdot \sin(\frac{k \gamma}{2}) \cdot \frac{1}{({ }_l \lambda_{n})^2 + ({ }_l \omega_{mnk}^{-})^2 } \left( \omega_{mnk}^{-} { }_l\Delta \alpha_{nm} - { }_l \lambda_{n} { }_l\Delta \beta_{nm} \right) \\
        { }_l P_{mnk}^{-s} =& -\frac{8}{c} \cdot \sin(\frac{k \gamma}{2}) \cdot \frac{1}{({ }_l \lambda_{n})^2 + ({ }_l \omega_{mnk}^{-})^2 } \left( \omega_{mnk}^{-} { }_l\Delta \alpha_{nm} + { }_l \lambda_{n} { }_l\Delta \beta_{nm} \right).
    \end{aligned}
\end{equation}

\subsection{Parameter estimation method}\label{subsection: parameter_estimation_method}

The parameters of Earth's free oscillations can be estimated using TianQin observed data and the analytical model derived above.
The frequencies and attenuation factors of various oscillation modes can be calculated using the \ac{PREM} model \cite{Dziewonski_Anderson_PREM_1981}, which is verified by ground experiments.
Meanwhile, the coefficients of intensities ${}_{l} \Delta \bar{C}_{n m}$ and ${}_{l} \Delta \bar{S}_{n m}$ depend on the physical processes of earthquakes \cite{stein_1977, Kanamori_1974, stein_2009_introduction}.
Therefore, we fix the frequencies and the attenuation factors of oscillation modes and estimate the coefficients of intensities to provide information for earthquake and related studies.

In this study, we adopt the Bayesian framework to obtain the posterior distribution of the parameters $\mathbf{\Theta}$ and to provide parameter estimation results and precision based on the posterior.
Bayesian theorem states that the posterior can be expressed as
\begin{equation}
    p(\mathbf{\Theta} \vert d, I) = \frac{p(d \vert \mathbf{\Theta}, I) p(\mathbf{\Theta} \vert I)}{p(d \vert I)},
\end{equation}
where $p(\mathbf{\Theta} \vert d, I)$, $p(d \vert \mathbf{\Theta}, I)$, $p(\mathbf{\Theta}\vert I)$ and $p(d \vert I)$ represent the posterior, the likelihood, the prior and the normalization factor, respectively.
The symbol $I$ represents the information about signal and noise.

For the TianQin mission, we usually assume that the noise in the detector is Gaussian and stationary, so the likelihood could be expressed as 
\begin{equation}
    \begin{aligned}
        \log \mathcal{L}(\mathbf{\Theta})= & \log p(d \vert \mathbf{\Theta}, I) \\
        = & -\frac{1}{2} \braket{ d-h(\mathbf{\Theta}) \vert d-h(\mathbf{\Theta})}+\text { const. } \\
        = & \braket{d \vert h(\mathbf{\Theta})} - \frac{1}{2} \braket{h(\mathbf{\Theta}) \vert h(\mathbf{\Theta})} - \frac{1}{2} \braket{d \vert d} + \text{ const.},
    \end{aligned}
\end{equation}
where $h(\mathbf{\Theta})$ denotes the waveform which we derive in the Section \ref{subsection: analytical_model_derivation}.
The symbol $\braket{g \vert h}$ represents the inner product between $g$ and $h$
\begin{equation}
    \braket{g \vert h} = 4 \mathcal{R} \int_{0}^{\infty} \frac{\tilde{g}(f) \cdot \tilde{h}^{*}(f)}{S_n(f)} \mathrm{d} f,
\end{equation}
where $S_n(f)$ denotes the one-side \ac{PSD} of the noise, and $\mathcal{R}$ represents the real component.

The constant terms in the likelihood can be absorbed into the normalization factor, rendering them negligible in our method for posterior calculation.
Consequently, the likelihood simplifies to
\begin{equation}
    \mathcal{L}(\mathbf{\Theta}) \propto \exp\left[\braket{d \vert h(\mathbf{\Theta})} - \frac{1}{2} \braket{h(\mathbf{\Theta}) \vert h(\mathbf{\Theta})}\right].
    \label{eq: waveform-likelihood}
\end{equation}

Within the Bayesian framework, the prior represents the information or knowledge prior to starting the analysis.
In our study, as we do not incorporate information from other observations, we assume a uniform distribution for the prior.
Subsequently, we proceed to calculate the posterior.

The Earth's free oscillations generated by a major earthquake, as analyzed in this study, contain up to eight observable modes for TianQin.
These modes contain 15 parameters.
Consequently, we must analyze a high-dimensional parameter space for posterior calculation.
Stochastic sampling methods, such as \ac{MCMC}, are commonly employed to efficiently explore the parameter space.
The MCMC method uses random walkers to generate samples, with a focus on moving towards regions of high posterior probability.
In this study, we employ \textbf{emcee}, an ensemble sampling \ac{MCMC} algorithm, for posterior distribution sampling \cite{Foreman-Mackey_Hogg_Lang_Goodman_2013}.
Subsequently, upon obtaining posterior distribution samples, properties of the source parameters like their means and variance can be derived (Fig. \ref{fig: geo-potential-coefficient-values} and \ref{fig: corner-with_true}).

However, since the \ac{MCMC} algorithm involves randomness during sampling, it is necessary to validate its estimation results using alternative methods.
In this study, the \ac{FIM} method is utilized for validation \cite{Cutler_1994_Gravitational,Vallisneri_2008_Use, Rodriguez_2013_Inadequacies}.
While estimating the parameters of \ac{GW}s, the \ac{FIM} can be defined as
\begin{equation}
    \Gamma_{ij} = \left \langle \frac{\partial h}{\partial \theta^i} \middle\vert \frac{\partial h}{\partial \theta^j} \right\rangle.
    \label{eq: fisher-information-matrix}
\end{equation}
Subsequently, the covariance matrix between all parameters can be approximated as $\mathbf{\Sigma} = \mathbf{\Gamma}^{-1}$.
Therefore, the estimation error can be estimated by
\begin{equation}
    \delta \theta^i = \sqrt{\Sigma^{ii}}.
\end{equation}
Comparing these results with the results obtained utilizing the \ac{MCMC} method allows us to verify the accuracy of the \ac{MCMC} parameter estimation.

\section{Results}\label{section: results}

\subsection{Analytical model verification}\label{subsection: analytical_model_verification}

\begin{figure}[h]%
    \centering
    \includegraphics[width=0.9\textwidth]{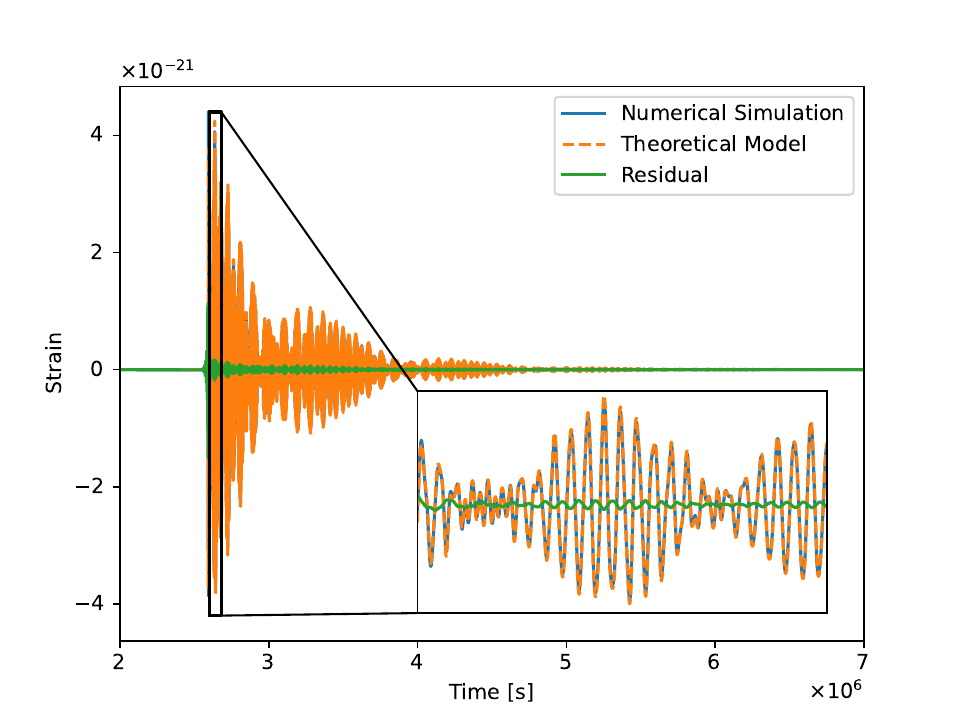}
    \caption{The figure illustrates the Earth's free oscillation response within the TianQin TDI-X channel, derived from both numerical simulation and analytical modeling approaches. The earthquake that induces the Earth's free oscillations occurs on day 30. The numerical simulation result is depicted by the blue solid line, while the analytical model result is represented by the orange dashed line. The green solid line illustrates the residual resulting from the subtraction of the aforementioned two results. With the exception of a minor region near the occurrence of the earthquake, the numerical result closely aligns with the predictions of the analytical formula.}
    \label{fig: x_channel_numerical_theoretica_compare}
\end{figure}

Prior to extracting Earth's free oscillation signals from TianQin simulation observation data, it is imperative to validate the efficacy of our analytical model.
Observation data are generated employing both the analytical waveform and numerical simulation methods, utilizing identical parameters derived from a magnitude 7.9 earthquake.
The resulting outputs from these divergent approaches are compared in Fig. \ref{fig: x_channel_numerical_theoretica_compare}.
It is evident that, apart from a minor discrepancy near the occurrence of the earthquake, the outcomes yielded by these two models exhibit remarkable proximity, signifying the efficacy of our analytical waveform in accurately describing Earth's free oscillations in TianQin.

\begin{figure}[h]%
    \centering
    \includegraphics[width=0.9\textwidth]{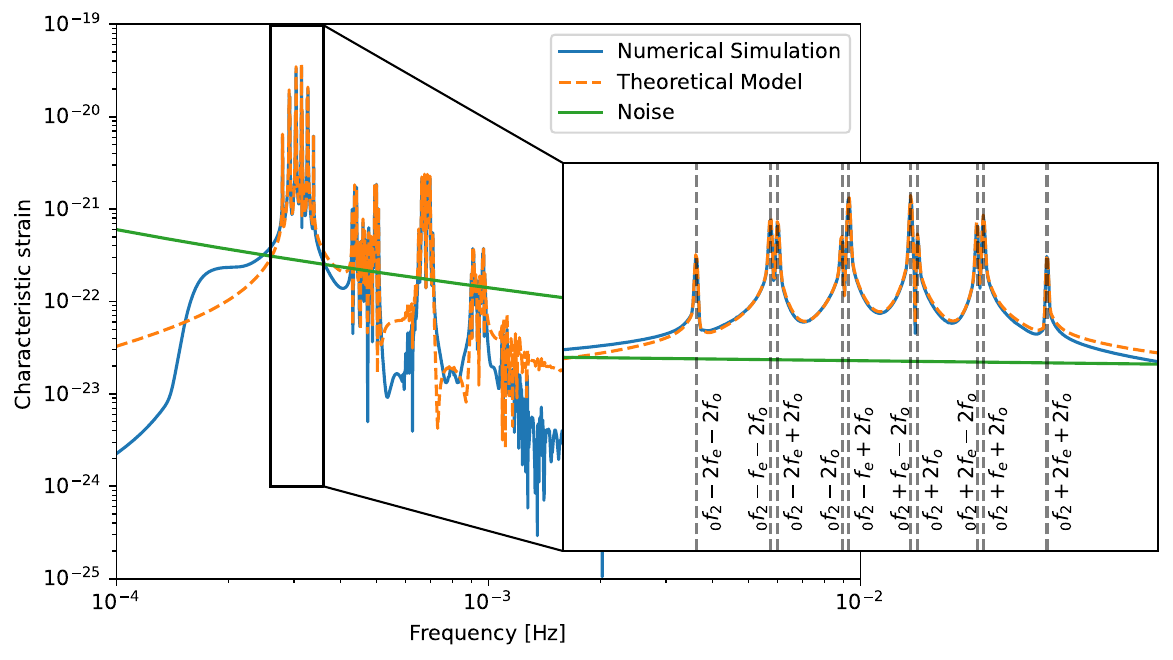}
    \caption{
        The figure illustrates the Earth's free oscillations response within the TianQin TDI-X channel in the frequency domain.
        Both numerical simulation data (blue solid line) and analytical modeling data (orange dashed line) are plotted in this figure.
        In the zoomed-in plot, the spectrum of the ${}_0 S_2$ mode is depicted.
        ${}_0 f_2$, $f_e$ and $f_o$ denote frequencies of ${}_0 S_2$ mode, Earth's rotation, and satellites' rotation respectively.
        It can be seen that this mode is split into 10 distinct spectral lines due to the influence of Earth's and satellites' rotation.
        Spectra splitting also occurs in other modes, and they can all be described by the analytical model (Eq. \eqref{eq: omega-relationship} \eqref{eq: TDI-X-channel-final-result} ).
        Different modes could add up deconstructive since they have opposite phases, making the overall peak appear to be slightly shifted.
        This is exactly what happens in certain frequencies like ${}_0 f_2 - f_e + 2f_o$, ${}_0 f_2 -f_e + 2f_o$, ${}_0 f_2 + f_e -2f_o$ and ${}_0 f_2 + f_e + 2f_o$.
        The green line represents the noise amplitude of TianQin TDI-X channel \cite{Li_2023_GWSpace}.
    }
    \label{fig: x_channel_frequency_domain}
\end{figure}

In Fig. \ref{fig: x_channel_frequency_domain}, we also show our waveform in the frequency domain, comparing it with the designed noise for TianQin.
Similar to the time domain result, the data are generated from the same magnitude 7.9 earthquake.
The characteristic strain \cite{Moore_2015_Gravitational_wave} in the frequency domain $    h_c(f) = f \vert \tilde{h}(f) \vert$ is shown on top of the noise $    h_n(f) = \sqrt{f S_n(f)}$, where $\tilde{h}(f)$ is the Fourier transform of the signal, and $S_n(f)$ represents the \ac{PSD} of noise.
One can define the \ac{SNR} of a signal as $\rho^2 = 4 \int_{0}^{\infty} \frac{\tilde{h}^*(f)\tilde{h}(f)}{S_n(f)} \mathrm{d} f$.
The \ac{SNR} of the signal shown in Fig. \ref{fig: x_channel_frequency_domain} is 73, confirming previous studies on the detectability. 

An important difference between space-based and ground-based detection can also be observed from Fig. \ref{fig: x_channel_frequency_domain}: the frequencies of Earth's normal modes have split due to the rotation of both the Earth and the satellites.
Our analytical waveform can predict the splitting frequency, which matches the simulation well.
The way frequencies split depends on the spherical harmonic function basis.
This allows us to directly estimate the different spherical harmonic amplitude of the Earth's normal modes, ${}_{l} \Delta \bar{C}_{n m}$ and ${}_{l}\Delta \bar{S}_{n m}$, using a single detector instead of joint detection with multiple detectors on the ground, which helps to avoid potential calibration errors among different detectors.

\subsection{Parameter estimation}\label{subsection: parameter_estimation}

In this section, we utilize the analytical waveform to extract signals from simulated observation data and estimate the parameters associated with Earth's normal modes.
The simulated observation data contains both numerical simulation signal from Earth's free oscillations generated by the Wenchuan earthquake and the background noise generated from TDI-X channel theoretical \ac{PSD} \cite{Luo_2016, Tinto_1999_Cancellation, Li_2023_GWSpace}.
The frequencies and attenuation factors of the Earth's free oscillations are held fixed, and we employ the \ac{MCMC} method to sample the posterior distribution of the amplitude coefficients ${}_{l} \Delta \bar{C}_{n m}$ and ${}_{l}\Delta \bar{S}_{n m}$.
Subsequently, we compute the means and variances of the sample set to obtain the estimates and accuracy.
To verify the accuracy of the parameter estimation results, we compare them with those obtained using the \ac{FIM} method \cite{Cutler_1994_Gravitational, Vallisneri_2008_Use, Rodriguez_2013_Inadequacies}.
For display purposes, we exclusively focus on the 9 modes with the highest response intensities, as the other high-order modes are submerged in background noise due to their weak response intensity.

\begin{figure}[h]%
    \centering
    \includegraphics[width=0.9\textwidth]{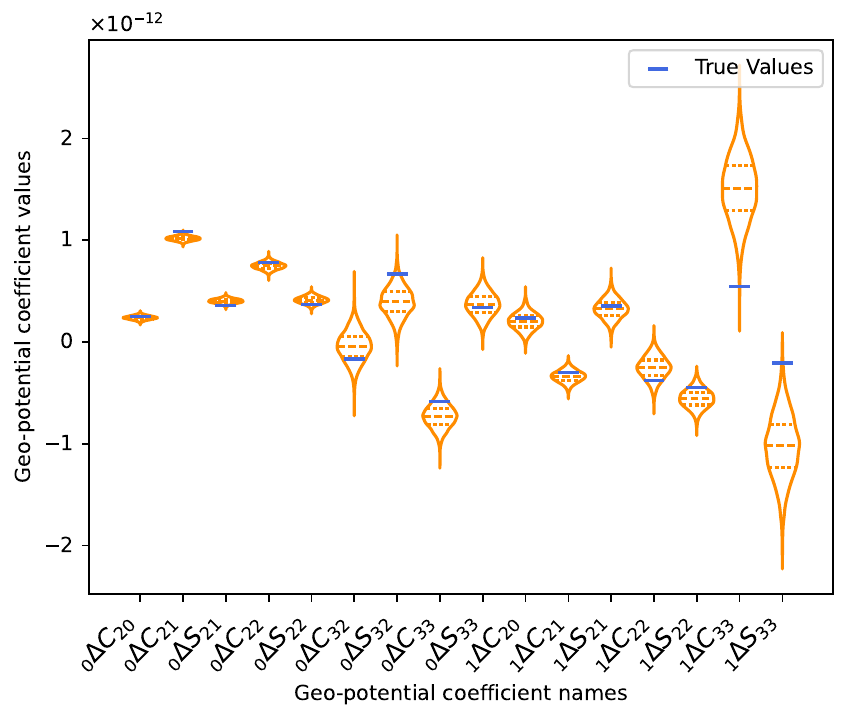}
    \caption{This figure shows the parameter estimation results and accuracy. The blue lines mark the true values we use to generate simulation data.}
    \label{fig: geo-potential-coefficient-values}
\end{figure}

\begin{figure}[h]%
    \centering
    \includegraphics[width=0.9\textwidth]{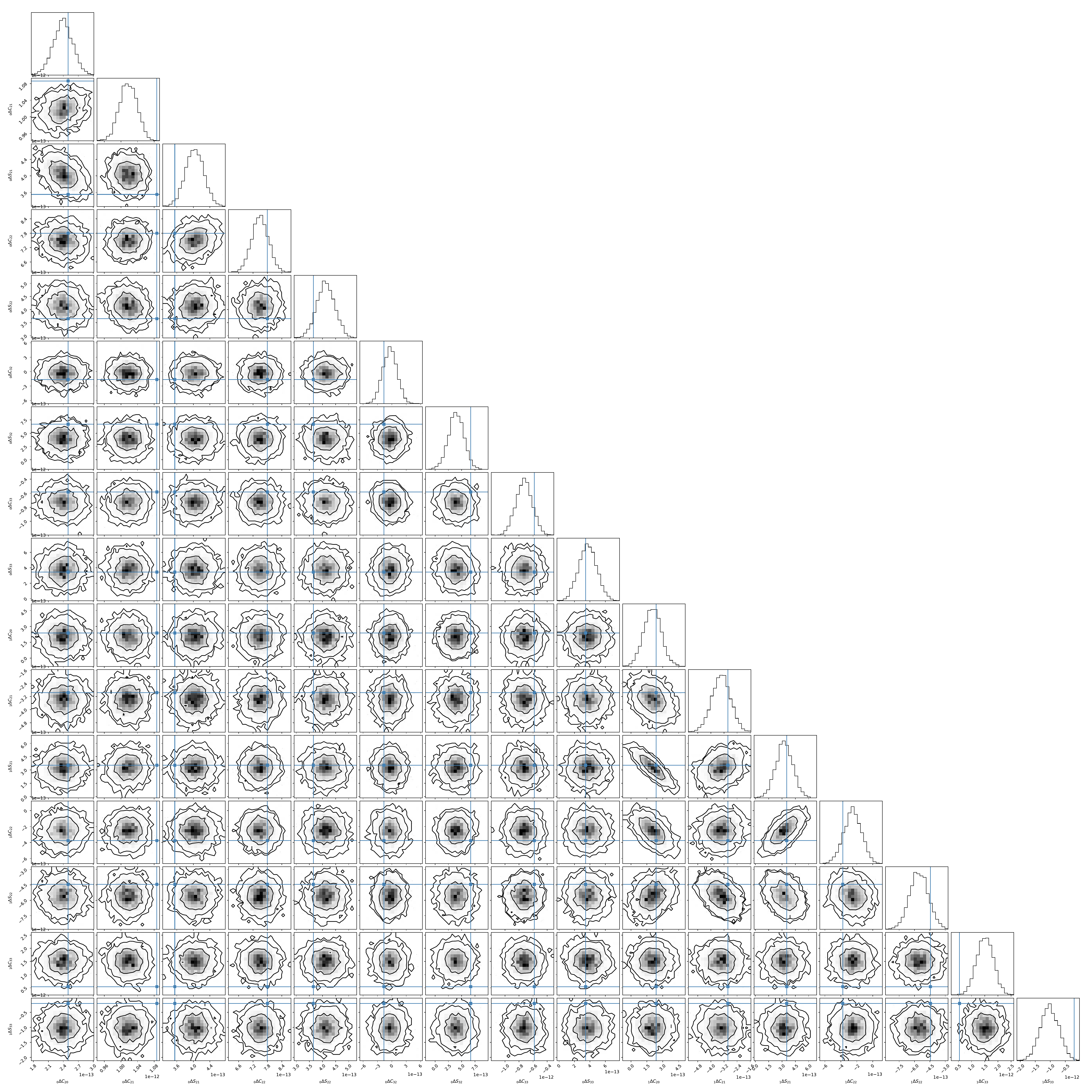}
    \caption{The corner plots show the two-dimensional posterior distribution between different parameters. The blue lines show the injected values of parameters in simulated data.}
    \label{fig: corner-with_true}
\end{figure}

The measured values of the parameters along with their errors are presented in Table \ref{tab: geopotential-coefficients}.
A more detailed and direct comparison can be made in Fig. \ref{fig: geo-potential-coefficient-values}, in which the violin plot is drawn for these parameters.
To avoid the impact of discrepancies between the analytical waveform and numerical simulation results near the occurrence of the earthquake, only the data from 5 hours to 15 days after the earthquake is applied for parameter estimation.

\begin{table}
    \centering
    \begin{tabular}{cccc}
        \hline
        Parameters & True values & Estimated values & FIM estimation errors \\
        \hline
        ${}_{0} \Delta \bar{C}_{2 0}$ & $2.496 \times 10^{-13}$ & $(2.39 \pm 0.20) \times 10^{-13}$ & $0.19 \times 10^{-13}$\\
        ${}_{0} \Delta \bar{S}_{2 0}$ & $0.0$ & - & - \\
        ${}_{0} \Delta \bar{C}_{2 1}$ & $1.086 \times 10^{-12}$ & $(1.018 \pm 0.021) \times 10^{-12}$ & $0.020 \times 10^{-12}$ \\
        ${}_{0} \Delta \bar{S}_{2 1}$ & $3.552 \times 10^{-13}$ & $(4.02 \pm 0.22) \times 10^{-13}$ & $0.23 \times 10^{-13}$ \\
        ${}_{0} \Delta \bar{C}_{2 2}$ & $7.799 \times 10^{-13}$ & $(7.49 \pm 0.35) \times 10^{-13}$ & $0.34 \times 10^{-13}$\\
        ${}_{0} \Delta \bar{S}_{2 2}$ & $3.650 \times 10^{-13}$ & $(4.13 \pm 0.35) \times 10^{-13}$ & $0.34 \times 10^{-13}$ \\
        ${}_{0} \Delta \bar{C}_{3 2}$ & $-1.662 \times 10^{-13}$ & $(-0.4 \pm 1.5) \times 10^{-13}$ & $1.4 \times 10^{-13}$ \\
        ${}_{0} \Delta \bar{S}_{3 2}$ & $6.689 \times 10^{-13}$ & $(4.0 \pm 1.5) \times 10^{-13}$ & $1.4 \times 10^{-13}$ \\
        ${}_{0} \Delta \bar{C}_{3 3}$ & $-5.813 \times 10^{-13}$ & $(-7.3 \pm 1.2) \times 10^{-13}$ & $1.2 \times 10^{-13}$ \\
        ${}_{0} \Delta \bar{S}_{3 3}$ & $3.436 \times 10^{-13}$ & $(3.7 \pm 1.2) \times 10^{-13}$ & $1.2 \times 10^{-13}$ \\
        ${}_{1} \Delta \bar{C}_{2 0}$ & $2.369 \times 10^{-13}$ & $(2.02 \pm 0.83) \times 10^{-13}$ & $0.60 \times 10^{-13}$ \\
        ${}_{1} \Delta \bar{S}_{2 0}$ & $0.0$ & - & - \\
        ${}_{1} \Delta \bar{C}_{2 1}$ & $-2.957 \times 10^{-13}$ & $(-3.37 \pm 0.58) \times 10^{-13}$ & $0.54 \times 10^{-13}$ \\
        ${}_{1} \Delta \bar{S}_{2 1}$ & $3.537 \times 10^{-13}$ & $(3.27 \pm 0.95) \times 10^{-13}$ & $0.77 \times 10^{-13}$ \\
        ${}_{1} \Delta \bar{C}_{2 2}$ & $-3.762 \times 10^{-13}$ & $(-2.5 \pm 1.1) \times 10^{-13}$ & $0.73 \times 10^{-13}$ \\
        ${}_{1} \Delta \bar{S}_{2 2}$ & $-4.451 \times 10^{-13}$ & $(-5.55 \pm 0.92) \times 10^{-13}$ & $0.82 \times 10^{-13}$ \\
        ${}_{1} \Delta \bar{C}_{3 3}$ & $5.431 \times 10^{-13}$ & $(15.1 \pm 3.2) \times 10^{-13}$ & $2.6 \times 10^{-13}$ \\
        ${}_{1} \Delta \bar{S}_{3 3}$ & $-2.053 \times 10^{-13}$ & $(-10.2 \pm 3.1) \times 10^{-13}$ & $2.4 \times 10^{-13}$ \\
        \hline
    \end{tabular}
    \caption{Corrections to the geopotential coefficients due to the earthquake and the estimated values. The last column shows the parameter estimation error obtained based on FIM, validating the accuracy of the parameter estimation results obtained by the MCMC method. }
    \label{tab: geopotential-coefficients}
\end{table}

\section{Discussion}\label{section: discussion}

This study introduces a novel approach to detecting Earth's free oscillations utilizing TianQin.
Unlike other space gravity measurement missions such as GRACE \cite{Ghobadi-Far_2019} and TianQin-2 \cite{Gong_2021_Concepts}, TianQin's satellites operate in high Earth orbit, mitigating the influence of Earth's high-order gravity field and facilitating the maintenance of quasi-circular orbits.
Crucially, the three assumptions -- quasi-circular orbits, equal orbital altitude, and an equilateral triangle constellation for the three satellites -- are fundamental for deriving analytical waveforms to characterize Earth's free oscillations in TianQin.
These assumptions allow the modulation effects of satellite motion on signals of Earth's free oscillations to be described using a single constant, the satellite orbital frequency, thereby neglecting high-order effects.
They also simplify the calculation for TDI channel.
Through a comparison with numerical simulation data in both the time domain and frequency domain, we observe a noteworthy concordance between the analytical waveform and the simulation data, affirming the efficacy of the analytical approach in capturing the response arising from Earth's free oscillations in TianQin.
The ability to formulate an analytical waveform for Earth's free oscillations represents a significant advantage afforded by high-orbit satellites, enabling the extraction of information from observation data beyond just observation.

Through the \ac{MCMC} method, we demonstrate that it is possible to extract physical information about the Earth's free oscillation from TianQin's observation data.
For the injected signal caused by the Wenchuan-like earthquake, TianQin is capable of detecting approximately 9 modes.
This result suggests that the \ac{GW} observatory, TianQin, can open a new window to examine the Earth's normal modes.
Compared to the ground-based stations, the satellite gravity method has a different source of noise, which can provide precious independent cross-validation.
Additionally, as shown in Eq. \ref{eq: variations-gravity-field-coefficients}, some of Earth's normal modes are frequency degeneracy, necessitating joint detection with multiple detectors for ground-based stations to distinguish them, which introduces potential calibration errors among different detectors.
The scanning effect of satellites can split these frequencies as Fig. \ref{fig: x_channel_frequency_domain} shown, thus breaking this degeneracy.
This allows TianQin to distinguish different modes independently, which can effectively avoid calibration errors.

The primary goal of the TianQin mission is to detect \ac{GW}s.
Earth's free oscillation signals and \ac{GW} signal may occur simultaneously on a rare occasion.
Due to the significant difference between the waveforms of these two kinds of signals, they are in principle unlikely to be confused.
When both signals are present in the observed data, a global analysis method can be employed to separate them \cite{Littenberg_2023}.
However, the details of method strategies require further investigation in the future.

Our analysis in this study focuses on the data in the TDI-X channel.
The symmetry of the TDI method \cite{Tinto_1999_Cancellation} allows for the straightforward generalization of our analytical waveform to the TDI-Y or TDI-Z channels by adjusting the phase of the satellite's orbit.
In the field of space GW detection, there is another widely used TDI channel known as the AET channel \cite{Armstrong_1999_Time}.
Unlike the XYZ channel, this channel consists of three orthogonal channels, with GW signals concentrated in the A and E channels, while the T channel primarily reflects detector noise.
Additionally, we conducted analyses using the AET channel.
Our findings suggest that our reconstructed waveform can effectively describe the response signals observed in both the TDI-A and TDI-E channels. 
Unlike the GW signals, numerical simulated data reveals high SNR responses of Earth's free oscillations in the TDI-T channel.
The reason for the different phenomena of Earth gravity perturbations and astrophysical-origin gravity perturbations in the TDI-T channel lies in the assumption that GWs can be approximated as plane waves, which holds true for the latter but fails for the former.
Similar to the result in \cite{Guo_2022_On}, the effectiveness of the plane wave assumption changes the way GWs respond, suggesting that the T channel no longer primarily reflects noise.
Unfortunately, the theoretical waveforms we currently derive cannot adequately describe the response of Earth's free oscillations in the TDI-T channel.
We plan to develop more sophisticated models to describe this response in the future, which will help separate the signals when TianQin observes both GWs and major earthquakes at the same time.

\section*{Acknowledgments}

\subsection*{General}

We thank Khosro Ghobadi-Far for providing helpful materials on GRACE's detection of Earth's free oscillations.
We thank Chengjian Luo and Lu Zheng for their support and assistance in generating numerical simulation data.
We thank Lei Jiao for his assistance with deriving the formulas.
We also thank Rongjiang Wang, Yong Zhang, Xiang'e Lei, Hao Zhou, Hang Li, Jiandong Zhang, Enkun Li, Zhengcheng Liang, and Zhiyuan Li for their helpful discussions and comments.

\subsection*{Author Contributions}

Yi-Ming Hu and Xuefeng Zhang designed the research. Yuxin Yang derived the formulas in this research. Yuxin Yang and Kun Liu analyzed the data. Yuxin Yang, Kun Liu, Xuefeng Zhang, and Yi-Ming Hu wrote the paper.

\subsection*{Funding}

X. Z. is supported by the National Key R\&D Program of China (Grant Nos. 2020YFC2201202 and 2022YFC2204600), NSFC (Grant No. 12373116), and Fundamental Research Funds for the Central Universities, Sun Yat-sen University (Grant No. 23xkjc001).
Y. H. is supported by the National Key Research and Development Program of China (No. 2020YFC2201400),  
the Natural Science Foundation of China (Grants  No.  12173104, No. 12261131504), and Guangdong Major Project of Basic and Applied Basic Research (Grant No. 2019B030302001).

\subsection*{Conflicts of Interest}

The authors declare that they have no conflict of interest.

\subsection*{Data Availability}

All data needed to evaluate the conclusions in the paper are present
in the paper and/or the Supplementary Materials. Additional data
related to this paper may be requested from the authors 

\section*{Supplementary Materials}

No additional supplementary materials were provided.

\printbibliography

@article{Zhang_2021,   title="{Effect of Earth-Moon's gravity on TianQin's range acceleration noise}",  volume={103},  url={http://dx.doi.org/10.1103/physrevd.103.062001},  DOI={10.1103/physrevd.103.062001},  number={6},  journal={Physical Review D},  author={Zhang, Xuefeng and Luo, Chengjian and Jiao, Lei and Ye, Bobing and Yuan, Huimin and Cai, Lin and Gu, Defeng and Mei, Jianwei and Luo, Jun},  year={2021},  month={Mar},  language={en-US}}

@article{Littenberg_2023,
    author = "Littenberg, Tyson B. and Cornish, Neil J.",
    title = "{Prototype global analysis of LISA data with multiple source types}",
    eprint = "2301.03673",
    archivePrefix = "arXiv",
    primaryClass = "gr-qc",
    doi = "10.1103/PhysRevD.107.063004",
    journal = "Phys. Rev. D",
    volume = "107",
    number = "6",
    pages = "063004",
    year = "2023"
}

@article{Jiao_2023,
  title = "{Effect of Earth-Moon's gravity on TianQin's range acceleration noise. III. An analytical model}",
  author = {Jiao, Lei and Zhang, Xuefeng},
  journal = {Phys. Rev. D},
  volume = {107},
  issue = {10},
  pages = {102004},
  numpages = {17},
  year = {2023},
  month = {May},
  publisher = {American Physical Society},
  doi = {10.1103/PhysRevD.107.102004},
  url = {https://link.aps.org/doi/10.1103/PhysRevD.107.102004}
}

@mastersthesis{Liu_2023_effects,
  author  = {Liu, Kun},
  title   = {Effects of Sun and Earth's free oscillations on TianQin's range acceleration noise},
  school  = {Sun Yat-sen University},
  year    = {2023}
}

@article{Luo_2016,   title="{TianQin: a space-borne gravitational wave detector}",  url={http://dx.doi.org/10.1088/0264-9381/33/3/035010},  DOI={10.1088/0264-9381/33/3/035010},  journal={Classical and Quantum Gravity},  author={Luo, Jun and Chen, Li-Sheng and Duan, Hui-Zong and Gong, Yun-Gui and Hu, Shoucun and Ji, Jianghui and Liu, Qi and Mei, Jianwei and Milyukov, Vadim and Sazhin, Mikhail and Shao, Cheng-Gang and Toth, Viktor T and Tu, Hai-Bo and Wang, Yamin and Wang, Yan and Yeh, Hsien-Chi and Zhan, Ming-Sheng and Zhang, Yonghe and Zharov, Vladimir and Zhou, Ze-Bing},  year={2016},  month={Feb},  pages={035010},  language={en-US}}

@article{Mei_2020_TianQin,
    author = {Mei, Jianwei and Bai, Yan-Zheng and Bao, Jiahui and Barausse, Enrico and Cai, Lin and Canuto, Enrico and Cao, Bin and Chen, Wei-Ming and Chen, Yu and Ding, Yan-Wei and Duan, Hui-Zong and Fan, Huimin and Feng, Wen-Fan and Fu, Honglin and Gao, Qing and Gao, TianQuan and Gong, Yungui and Gou, Xingyu and Gu, Chao-Zheng and Gu, De-Feng and He, Zi-Qi and Hendry, Martin and Hong, Wei and Hu, Xin-Chun and Hu, Yi-Ming and Hu, Yuexin and Huang, Shun-Jia and Huang, Xiang-Qing and Jiang, Qinghua and Jiang, Yuan-Ze and Jiang, Yun and Jiang, Zhen and Jin, Hong-Ming and Korol, Valeriya and Li, Hong-Yin and Li, Ming and Li, Ming and Li, Pengcheng and Li, Rongwang and Li, Yuqiang and Li, Zhu and Li, Zhulian and Li, Zhu-Xi and Liang, Yu-Rong and Liang, Zheng-Cheng and Liao, Fang-Jie and Liu, Qi and Liu, Shuai and Liu, Yan-Chong and Liu, Li and Liu, Pei-Bo and Liu, Xuhui and Liu, Yuan and Lu, Xiong-Fei and Lu, Yang and Lu, Ze-Huang and Luo, Yan and Luo, Zhi-Cai and Milyukov, Vadim and Ming, Min and Pi, Xiaoyu and Qin, Chenggang and Qu, Shao-Bo and Sesana, Alberto and Shao, Chenggang and Shi, Changfu and Su, Wei and Tan, Ding-Yin and Tan, Yujie and Tan, Zhuangbin and Tu, Liang-Cheng and Wang, Bin and Wang, Cheng-Rui and Wang, Fengbin and Wang, Guan-Fang and Wang, Haitian and Wang, Jian and Wang, Lijiao and Wang, Panpan and Wang, Xudong and Wang, Yan and Wang, Yi-Fan and Wei, Ran and Wu, Shu-Chao and Xiao, Chun-Yu and Xu, Xiao-Shi and Xue, Chao and Yang, Fang-Chao and Yang, Liang and Yang, Ming-Lin and Yang, Shan-Qing and Ye, Bobing and Yeh, Hsien-Chi and Yu, Shenghua and Zhai, Dongsheng and Zhang, Caishi and Zhang, Haitao and Zhang, Jian-dong and Zhang, Jie and Zhang, Lihua and Zhang, Xin and Zhang, Xuefeng and Zhou, Hao and Zhou, Ming-Yue and Zhou, Ze-Bing and Zhu, Dong-Dong and Zi, Tie-Guang and Luo, Jun},
    title = "{The TianQin project: Current progress on science and technology}",
    journal = {Progress of Theoretical and Experimental Physics},
    volume = {2021},
    number = {5},
    pages = {05A107},
    year = {2020},
    month = {08},
    issn = {2050-3911},
    doi = {10.1093/ptep/ptaa114},
    url = {https://doi.org/10.1093/ptep/ptaa114},
    eprint = {https://academic.oup.com/ptep/article-pdf/2021/5/05A107/37953035/ptaa114.pdf},
}

@article{Gong_2021_Concepts,
  title="{Concepts and status of Chinese space gravitational wave detection projects}",
  author={Gong, Yungui and Luo, Jun and Wang, Bin},
  journal={Nature Astronomy},
  volume={5},
  number={9},
  pages={881--889},
  year={2021},
  publisher={Nature Publishing Group UK London},
  doi={10.1038/s41550-021-01480-3},
  url={https://doi.org/10.1038/s41550-021-01480-3},
}

@article{Li_2023_GWSpace,
    author = "Li, En-Kun and others",
    title = "{GWSpace: a multi-mission science data simulator for space-based gravitational wave detection}",
    eprint = "2309.15020",
    archivePrefix = "arXiv",
    primaryClass = "gr-qc",
    month = "9",
    year = "2023"
}

@article{Tinto_Dhurandhar_2014,   title="{Time-Delay Interferometry}",  url={http://dx.doi.org/10.12942/lrr-2014-6},  DOI={10.12942/lrr-2014-6},  journal={Living Reviews in Relativity},  author={Tinto, Massimo and Dhurandhar, Sanjeev V.},  year={2014},  month={Dec},  language={en-US}  }

@article{Armstrong_1999_Time,
  title={Time-delay interferometry for space-based gravitational wave searches},
  author={Armstrong, JW and Estabrook, FB and Tinto, Massimo},
  journal={The Astrophysical Journal},
  volume={527},
  number={2},
  pages={814},
  year={1999},
  publisher={IOP Publishing}
}

@article{Tinto_1999_Cancellation,
  title={Cancellation of laser noise in an unequal-arm interferometer detector of gravitational radiation},
  author={Tinto, Massimo and Armstrong, John W},
  journal={Physical Review D},
  volume={59},
  number={10},
  pages={102003},
  year={1999},
  publisher={APS}
}

@article{Foreman-Mackey_Hogg_Lang_Goodman_2013,  
 title="{emcee: The MCMC Hammer}", 
 url={http://dx.doi.org/10.1086/670067}, 
 DOI={10.1086/670067}, 
 journal={Publications of the Astronomical Society of the Pacific}, 
 author={Foreman-Mackey, Daniel and Hogg, David W. and Lang, Dustin and Goodman, Jonathan}, 
 year={2013}, 
 month={Mar}, 
 pages={306–312}, 
 language={en-US} 
 }

@article{Ghobadi-Far_2019,  
 title="{Gravitational Changes of the Earth's Free Oscillation From Earthquakes: Theory and Feasibility Study Using GRACE Inter-satellite Tracking}", 
 volume={124}, 
 url={http://dx.doi.org/10.1029/2019jb017530}, 
 DOI={10.1029/2019jb017530}, 
 number={7}, 
 journal={Journal of Geophysical Research: Solid Earth}, 
 author={Ghobadi-Far, Khosro and Han, Shin-Chan and Sauber, Jeanne and Lemoine, Frank and Behzadpour, Saniya and Mayer-Gürr, Torsten and Sneeuw, Nico and Okal, Emile}, 
 year={2019}, 
 month={Jul}, 
 pages={7483-7503}, 
 language={en-US} 
 }

@article{Han_Riva_Sauber_Okal_2013,  
 title={Source parameter inversion for recent great earthquakes from a decade-long observation of global gravity fields}, 
 volume={118}, 
 url={http://dx.doi.org/10.1002/jgrb.50116}, 
 DOI={10.1002/jgrb.50116}, 
 number={3}, 
 journal={Journal of Geophysical Research: Solid Earth}, 
 author={Han, Shin-Chan and Riva, Riccardo and SauPer, Jeanne and Okal, Emile}, 
 year={2013}, 
 month={Mar}, 
 pages={1240-1267}, 
 language={en-US} 
 }

@article{Tapley_Bettadpur_Watkins_Reigber_GRACE_2004,  
 title={The gravity recovery and climate experiment: Mission overview and early results}, 
 url={http://dx.doi.org/10.1029/2004gl019920}, 
 DOI={10.1029/2004gl019920}, 
 journal={Geophysical Research Letters}, 
 author={Tapley, B. D. and Bettadpur, S. and Watkins, M. and Reigber, C.}, 
 year={2004}, 
 month={May}, 
 pages={n/a-n/a}, 
 language={en-US} 
 }

@article{Kornfeld_Arnold_Gross_Dahya_Klipstein_Gath_Bettadpur_GRACE-FO_2019,  
 title="{GRACE-FO: The Gravity Recovery and Climate Experiment Follow-On Mission}", 
 volume={56}, 
 url={http://dx.doi.org/10.2514/1.a34326}, 
 DOI={10.2514/1.a34326}, 
 number={3}, 
 journal={Journal of Spacecraft and Rockets}, 
 author={Kornfeld, Richard P. and Arnold, Bradford W. and Gross, Michael A. and Dahya, Neil T. and Klipstein, William M. and Gath, Peter F. and Bettadpur, Srinivas}, 
 year={2019}, 
 month={May}, 
 pages={931–951}, 
 language={en-US} 
 }

@article{Abich_2019_In-Orbit,
  title = {In-Orbit Performance of the GRACE Follow-on Laser Ranging Interferometer},
  author = {Abich, Klaus and Abramovici, Alexander and Amparan, Bengie and Baatzsch, Andreas and Okihiro, Brian Bachman and Barr, David C. and Bize, Maxime P. and Bogan, Christina and Braxmaier, Claus and Burke, Michael J. and Clark, Ken C. and Dahl, Christian and Dahl, Katrin and Danzmann, Karsten and Davis, Mike A. and de Vine, Glenn and Dickson, Jeffrey A. and Dubovitsky, Serge and Eckardt, Andreas and Ester, Thomas and Barranco, Germ\'an Fern\'andez and Flatscher, Reinhold and Flechtner, Frank and Folkner, William M. and Francis, Samuel and Gilbert, Martin S. and Gilles, Frank and Gohlke, Martin and Grossard, Nicolas and Guenther, Burghardt and Hager, Philipp and Hauden, Jerome and Heine, Frank and Heinzel, Gerhard and Herding, Mark and Hinz, Martin and Howell, James and Katsumura, Mark and Kaufer, Marina and Klipstein, William and Koch, Alexander and Kruger, Micah and Larsen, Kameron and Lebeda, Anton and Lebeda, Arnold and Leikert, Thomas and Liebe, Carl Christian and Liu, Jehhal and Lobmeyer, Lynette and Mahrdt, Christoph and Mangoldt, Thomas and McKenzie, Kirk and Misfeldt, Malte and Morton, Phillip R. and M\"uller, Vitali and Murray, Alexander T. and Nguyen, Don J. and Nicklaus, Kolja and Pierce, Robert and Ravich, Joshua A. and Reavis, Gretchen and Reiche, Jens and Sanjuan, Josep and Sch\"utze, Daniel and Seiter, Christoph and Shaddock, Daniel and Sheard, Benjamin and Sileo, Michael and Spero, Robert and Spiers, Gary and Stede, Gunnar and Stephens, Michelle and Sutton, Andrew and Trinh, Joseph and Voss, Kai and Wang, Duo and Wang, Rabi T. and Ware, Brent and Wegener, Henry and Windisch, Steve and Woodruff, Christopher and Zender, Bernd and Zimmermann, Marcus},
  journal = {Phys. Rev. Lett.},
  volume = {123},
  issue = {3},
  pages = {031101},
  numpages = {7},
  year = {2019},
  month = {Jul},
  publisher = {American Physical Society},
  doi = {10.1103/PhysRevLett.123.031101},
  url = {https://link.aps.org/doi/10.1103/PhysRevLett.123.031101}
}

@article{Dziewonski_Anderson_PREM_1981,  
 title="{Preliminary reference Earth model}", 
 url={http://dx.doi.org/10.1016/0031-9201(81)90046-7}, 
 DOI={10.1016/0031-9201(81)90046-7}, 
 journal={Physics of the Earth and Planetary Interiors}, 
 author={Dziewonski, Adam M. and Anderson, Don L.}, 
 year={1981}, 
 month={Jun}, 
 pages={297-356}, 
 language={en-US} 
 }

@Book{Montenbruck_2000,
  author    = {O. Montenbruck and E. Gill},
  publisher = {Springer},
  title     = "{Satellite Orbits - Models, Methods, and Applications}",
  year      = {2000},
}

@article{Kaula_1961,  
 title="{Analysis of Gravitational and Geometric Aspects of Geodetic Utilization of Satellites}", 
 volume={5}, 
 url={http://dx.doi.org/10.1111/j.1365-246x.1961.tb00417.x}, 
 DOI={10.1111/j.1365-246x.1961.tb00417.x}, 
 number={2}, 
 journal={Geophysical Journal International}, 
 author={Kaula, W. M.}, 
 pages={104-133}, 
 language={en-US},
 year={1961}
 }

@Book{Kaula_1966,
  author    = {W. M. Kaula},
  publisher = {Blaisdell Publishing Company},
  title     = "{Theory of Satellite Geodesy}",
  year      = {1966},
}

@PhdThesis{Sharifi_2006,
  author = {M. A. Sharifi},
  title  = "{Satellite to Satellite Tracking in the Space-wise Approach}",
  school = {University of Stuttgart},
  year   = {2006},
  type   = {{Ph.D.} thesis},
}

@book{stein_2009_introduction,
  title={An introduction to seismology, earthquakes, and earth structure},
  author={Stein, Seth and Wysession, Michael},
  year={2009},
  publisher={John Wiley \& Sons}
}

@article{stein_1977,
  title="{Amplitudes of the Earth's split normal modes}",
  author={Stein, Seth and Geller, Robert J},
  journal={Journal of Physics of the Earth},
  volume={25},
  number={2},
  pages={117--142},
  year={1977},
  publisher={The Seismological Society of Japan, The Volcanological Society of Japan, The~…}
}

@article{Kanamori_1974,
title = "{Focal process of the great Chilean earthquake May 22, 1960}",
journal = {Physics of the Earth and Planetary Interiors},
volume = {9},
number = {2},
pages = {128-136},
year = {1974},
issn = {0031-9201},
doi = {https://doi.org/10.1016/0031-9201(74)90029-6},
url = {https://www.sciencedirect.com/science/article/pii/0031920174900296},
author = {Hiroo Kanamori and John J. Cipar}
}

@article{Ness_1961_observations,
  title={Observations of the free oscillations of the earth},
  author={Ness, NF and Harrison, JC and Slichter, LB},
  journal={Journal of Geophysical Research},
  volume={66},
  number={2},
  pages={621--629},
  year={1961},
  publisher={Wiley Online Library}
}

@article{Benioff_1961_excitation,
  title="{Excitation of the free oscillations of the Earth by earthquakes}",
  author={Benioff, Hugo and Press, Frank and Smith, Stewart},
  journal={Journal of Geophysical Research},
  volume={66},
  number={2},
  pages={605--619},
  year={1961},
  publisher={Wiley Online Library}
}

@article{Dziewonski_1971_Overtones,
  title={Overtones of free oscillations and the structure of the Earth's interior},
  author={Dziewonski, Adam M},
  journal={Science},
  volume={172},
  number={3990},
  pages={1336--1338},
  year={1971},
  publisher={American Association for the Advancement of Science}
}

@article{Dziewonski_1972_Observations,
  title="{Observations of normal modes from 84 recordings of the Alaskan earthquake of 1964 March 28}",
  author={Dziewonski, AM and Gilbert, Freeman},
  journal={Geophysical Journal International},
  volume={27},
  number={4},
  pages={393--446},
  year={1972},
  publisher={Blackwell Publishing Ltd Oxford, UK}
}

@article{Geller_1979_time,
  title={Time-domain attenuation measurements for fundamental spheroidal modes {(${}_{0}S_{6}$ to ${}_{0}S_{28}$)} for the 1977 {Indonesian} earthquake},
  author={Geller, Robert J and Stein, Seth},
  journal={Bulletin of the Seismological Society of America},
  volume={69},
  number={6},
  pages={1671--1691},
  year={1979},
  publisher={The Seismological Society of America}
}

@article{Buland_1979_Observations,
  title="{Observations from the IDA network of attenuation and splitting during a recent earthquake}",
  author={Buland, R and Berger, J and Gilbert, Freeman},
  journal={Nature},
  volume={277},
  number={5695},
  pages={358--362},
  year={1979},
  publisher={Nature Publishing Group UK London}
}

@article{Ding_2013_Search,
  title="{Search for the Slichter modes based on a new method: Optimal sequence estimation}",
  author={Ding, Hao and Shen, Wen-Bin},
  journal={Journal of Geophysical Research: Solid Earth},
  volume={118},
  number={9},
  pages={5018--5029},
  year={2013},
  publisher={Wiley Online Library}
}

@article{Courtier_2000_Global,  
 title={Global superconducting gravimeter observations and the search for the translational modes of the inner core}, 
 volume={117}, 
 url={http://dx.doi.org/10.1016/s0031-9201(99)00083-7}, 
 DOI={10.1016/s0031-9201(99)00083-7}, 
 number={1–4}, 
 journal={Physics of the Earth and Planetary Interiors}, 
 author={Courtier, N. and Ducarme, B. and Goodkind, J. and Hinderer, J. and Imanishi, Y. and Seama, N. and Sun, H. and Merriam, J. and Bengert, B. and Smylie, D.E.}, 
 year={2000}, 
 month={Jan}, 
 pages={3–20}, 
 language={en-US} 
 }

@article{Rosat_2005_high,
  title={High-resolution analysis of the gravest seismic normal modes after the 2004 {$M_w=9$} {Sumatra} earthquake using superconducting gravimeter data},
  author={Rosat, Severine and Sato, Tadahiro and Imanishi, Yuichi and Hinderer, J and Tamura, Yoshiaki and McQueen, Herbert and Ohashi, M},
  journal={Geophysical research letters},
  volume={32},
  number={13},
  year={2005},
  publisher={Wiley Online Library}
}

@article{Ding_2015_data,
  title="{Data stacking methods for isolation of normal-mode singlets of Earth's free oscillation: Extensions, comparisons, and applications}",
  author={Ding, Hao and Chao, Benjamin F},
  journal={Journal of Geophysical Research: Solid Earth},
  volume={120},
  number={7},
  pages={5034--5050},
  year={2015},
  publisher={Wiley Online Library}
}

@article{Luan_2021_Progress,
  title="{Progress and prospect of studies on elastic normal modes of Earth's free oscillation}",
  author={Luan, W and Shen, WB and Ding, H},
  journal={Reviews of Geophysics and Planetary Physics},
  volume={52},
  number={3},
  pages={308--325},
  year={2021}
}

@article{Woodhouse_2010_theory,
  title={Theory and observations--Earth's free oscillations},
  author={Woodhouse, JH and Deuss, A and Schubert, G},
  journal={Seismology and Structure of the Earth: Treatise on Geophysics},
  volume={1},
  pages={31},
  year={2010},
  publisher={Elsevier}
}

@article{Park_2005_earth,
  title="{Earth's free oscillations excited by the 26 December 2004 Sumatra-Andaman earthquake}",
  author={Park, Jeffrey and Song, Teh-Ru Alex and Tromp, Jeroen and Okal, Emile and Stein, Seth and Roult, Genevieve and Clevede, Eric and Laske, Gabi and Kanamori, Hiroo and Davis, Peter and others},
  journal={Science},
  volume={308},
  number={5725},
  pages={1139--1144},
  year={2005},
  publisher={American Association for the Advancement of Science}
}

@article{Gilbert_Dziewonski_1975,  
 title={An application of normal mode theory to the retrieval of structural parameters and source mechanisms from seismic spectra}, 
 volume={278}, 
 url={http://dx.doi.org/10.1098/rsta.1975.0025}, 
 DOI={10.1098/rsta.1975.0025}, 
 number={1280}, 
 journal={Philosophical Transactions of the Royal Society of London. Series A, Mathematical and Physical Sciences}, 
 author={Gilbert, Freeman and Dziewonski, AdamM.}, 
 year={1975}, 
 month={Mar}, 
 pages={187–269}, 
 language={en-US} 
 }

@article{Masters_Jordan_Silver_Gilbert_1982,  
 title="{Aspherical Earth structure from fundamental spheroidal-mode data}", 
 volume={298}, 
 url={http://dx.doi.org/10.1038/298609a0}, 
 DOI={10.1038/298609a0}, 
 number={5875}, 
 journal={Nature}, 
 author={Masters, Guy and Jordan, Thomas H. and Silver, Paul G. and Gilbert, Freeman}, 
 year={1982}, 
 month={Aug}, 
 pages={609–613}, 
 language={en-US} 
 }

@article{Ritzwoller_Lavely_1995,  
 title="{Three-dimensional seismic models of the Earth’s mantle}", 
 volume={33}, 
 url={http://dx.doi.org/10.1029/94rg03020}, 
 DOI={10.1029/94rg03020}, 
 number={1}, 
 journal={Reviews of Geophysics}, 
 author={Ritzwoller, Michael H. and Lavely, Eugene M.}, 
 year={1995}, 
 month={Feb}, 
 pages={1–66}, 
 language={en-US} 
 }

@article{Romanowicz_2000_anomalous,
  title="{Anomalous splitting of free oscillations: A reevaluation of possible interpretations}",
  author={Romanowicz, Barbara and Br{\'e}ger, Ludovic},
  journal={Journal of Geophysical Research: Solid Earth},
  volume={105},
  number={B9},
  pages={21559--21578},
  year={2000},
  publisher={Wiley Online Library}
}

@article{Dziewonski_1971_solidity,
  title="{Solidity of the inner core of the Earth inferred from normal mode observations}",
  author={Dziewonski, AM and Gilbert, F},
  journal={Nature},
  volume={234},
  number={5330},
  pages={465--466},
  year={1971},
  publisher={Nature Publishing Group UK London}
}

@article{Woodhouse_1986_evidence,
  title={Evidence for inner core anisotropy from free oscillations},
  author={Woodhouse, John H and Giardini, Domenico and Li, Xiang-Dong},
  journal={Geophysical Research Letters},
  volume={13},
  number={13},
  pages={1549--1552},
  year={1986},
  publisher={Wiley Online Library}
}

@article{Tromp_1993_support,
  title="{Support for anisotropy of the Earth's inner core from free oscillations}",
  author={Tromp, Jeroen},
  journal={Nature},
  volume={366},
  number={6456},
  pages={678--681},
  year={1993},
  publisher={Nature Publishing Group UK London}
}

@article{Ishii_1999_Normal,
  title={Normal-mode and free-air gravity constraints on lateral variations in velocity and density of Earth's mantle},
  author={Ishii, Miaki and Tromp, Jeroen},
  journal={Science},
  volume={285},
  number={5431},
  pages={1231--1236},
  year={1999},
  publisher={American Association for the Advancement of Science}
}

@article{Rosat_2003_First,
  title={First observation of 2S1 and study of the splitting of the football mode 0S2 after the June 2001 Peru earthquake of magnitude 8.4},
  author={Rosat, Severine and Hinderer, Jacques and Rivera, Luis},
  journal={Geophysical Research Letters},
  volume={30},
  number={21},
  year={2003},
  publisher={Wiley Online Library}
}

@article{Deuss_2013_New,
  title={A new catalogue of normal-mode splitting function measurements up to 10 mHz},
  author={Deuss, Arwen and Ritsema, Jeroen and van Heijst, Hendrik},
  journal={Geophysical Journal International},
  volume={193},
  number={2},
  pages={920--937},
  year={2013},
  publisher={Oxford University Press}
}

@article{Guo_2022_On,
doi = {10.3847/1538-4357/ac9131},
url = {https://dx.doi.org/10.3847/1538-4357/ac9131},
year = {2022},
month = {nov},
publisher = {The American Astronomical Society},
volume = {939},
number = {1},
pages = {55},
author = {Xiao Guo and Youjun Lu and Qingjuan Yu},
title = "{On Detecting Nearby Nanohertz Gravitational Wave Sources via Pulsar Timing Arrays}",
journal = {The Astrophysical Journal}
}

@article{Christensen_2022_Parameter,
  title={Parameter estimation with gravitational waves},
  author={Christensen, Nelson and Meyer, Renate},
  journal={Reviews of modern physics},
  volume={94},
  number={2},
  pages={025001},
  year={2022},
  publisher={APS}
}

@article{Cutler_1994_Gravitational,
  title="{Gravitational waves from merging compact binaries: How accurately can one extract the binary's parameters from the inspiral waveform?}",
  author={Cutler, Curt and Flanagan, Eanna E},
  journal={Physical Review D},
  volume={49},
  number={6},
  pages={2658},
  year={1994},
  publisher={APS}
}

@article{Vallisneri_2008_Use,
  title={Use and abuse of the Fisher information matrix in the assessment of gravitational-wave parameter-estimation prospects},
  author={Vallisneri, Michele},
  journal={Physical Review D},
  volume={77},
  number={4},
  pages={042001},
  year={2008},
  publisher={APS}
}

@article{Rodriguez_2013_Inadequacies,
  title={Inadequacies of the Fisher information matrix in gravitational-wave parameter estimation},
  author={Rodriguez, Carl L and Farr, Benjamin and Farr, Will M and Mandel, Ilya},
  journal={Physical Review D},
  volume={88},
  number={8},
  pages={084013},
  year={2013},
  publisher={APS}
}

@article{Moore_2015_Gravitational_wave,
doi = {10.1088/0264-9381/32/1/015014},
url = {https://dx.doi.org/10.1088/0264-9381/32/1/015014},
year = {2014},
month = {dec},
publisher = {IOP Publishing},
volume = {32},
number = {1},
pages = {015014},
author = {C J Moore and R H Cole and C P L Berry},
title = {Gravitational-wave sensitivity curves},
journal = {Classical and Quantum Gravity}
}

\end{document}